\documentclass{article}
\usepackage{ijcai20}

\usepackage{times}

\usepackage{soul}
\usepackage{url}
\usepackage[utf8]{inputenc}
\usepackage[small]{caption}
\usepackage{amsmath}
\usepackage{siunitx}
\usepackage{booktabs}
\usepackage{cite}
\usepackage{caption}
\usepackage{multirow}
\usepackage{flushend}

\usepackage[ruled, vlined]{algorithm2e}

\usepackage[none]{hyphenat}
\usepackage{geometry}
\newif\ifnotes
\notestrue 

\usepackage{graphicx}
\usepackage[dvipsnames]{xcolor}
\usepackage[
  colorlinks=true,
  linkcolor=purple,
  citecolor=purple,
  urlcolor=purple,
  pdfauthor={},
  pdftitle={},
  pdfsubject={},
  pdfkeywords={},
  bookmarks=false,
]{hyperref}

\usepackage{cleveref}
\crefformat{section}{\S#2#1#3}
\crefformat{subsection}{\S#2#1#3}
\crefformat{subsubsection}{\S#2#1#3}
\crefrangeformat{section}{\S\S#3#1#4 to~#5#2#6}
\crefmultiformat{section}{\S\S#2#1#3}{ and~#2#1#3}{, #2#1#3}{ and~#2#1#3}
\crefname{algorithm}{Algorithm}{Algorithms}
\crefname{figure}{Figure}{Figures}
\crefname{appendix}{Appendix}{Appendix}
\crefname{table}{Table}{Tables}

\title{Optimizing Streamlined Blockchain Consensus with Generalized Weighted Voting and Enhanced Leader Rotation\thanks{This work has been done in the context of a Bachelor research project, see \url{https://resolver.tudelft.nl/uuid:907bd47c-394a-4e26-b901-de713159dcb8}.}}

\author{Diana Micloiu \and Rowdy Chotkan \and Jérémie Decouchant\\
Delft University of Technology\\
The Netherlands\\
\{D.Micloiu, R.M.Chotkan-1, j.decouchant\}@tudelft.nl
}

\begin{document}

\maketitle

\thispagestyle{plain}
\pagestyle{plain}

\begin{abstract}
\textit{Streamlined} Byzantine Fault Tolerant (BFT) protocols, such as HotStuff~[PODC'19], and \textit{weighted voting} represent two possible strategies to improve consensus in the distributed systems world. Several studies have been conducted on both techniques, but the research on combining the two is scarce. To cover this knowledge gap, we introduce a weighted voting approach on Hotstuff, along with two optimisations targeting weight assignment distribution and leader rotation in the underlying state replication protocol. Moreover, the weighted protocols developed rely on studies proving the effectiveness of a specific voting power assignment based on discrete values. We generalise this approach by presenting a novel continuous weighting scheme applied to the Hotstuff protocol to highlight the effectiveness of this technique in faulty scenarios. We prove the significant latency reduction impact of weighted voting on streamlined protocols and advocate for further research.
\end{abstract} 
\section{Introduction} 

\looseness=-1 Many distributed system paradigms, such as state machine replication~\cite{SMR} and blockchains, have a common core concept: \textit{consensus}, which denotes the collective agreement of network participants. Distributed systems are known to be prone to hardware and software failures that can compromise availability or even change the system's normal behaviour. Hence, consensus is needed as a mechanism for coordinating the system's critical actions and ensuring its functionality.

\looseness=-1 In the blockchain world, consensus algorithms lay at the basis of distributed ledger technologies. They are classified into permissioned and permissionless, distinguishing each other by either limiting participation to a predetermined set of nodes or allowing anyone to join. Out of the two, permissionless systems gained more popularity, with the seminal Nakamoto consensus relying on Proof-of-Work. However, its significant impact on energy consumption revealed the system's limitations and urged researchers to look for alternative consensus algorithms~\cite{engcons}. In turn, interest in permissioned systems grew as their efficiency in terms of throughput, latency, and finality was observed. Thus, the focus shifted towards finding ways to optimise their performance. Strategies such as system size reduction~\cite{ThreatAdapt} and leader rotation mechanisms~\cite{amir2010prime} have been explored to enhance scalability and resilience. Notably, the Practical Byzantine Fault Tolerance algorithm (PBFT)~\cite{PBFT} has been a focal point of research in permissioned systems.

\looseness=-1 PBFT is part of a more prominent family of protocols: Byzantine Fault Tolerant (BFT), which enables systems to tolerate arbitrary node failures~\cite{lamport2019byzantine , cachin2017blockchain , natoli2019deconstructing}. In particular, the protocol requires $3f + 1$ nodes in the system to withstand $f$ failures. Hence, the capability of the system to resist failures comes with the cost of managing the demand for the increased number of nodes and higher communication complexity. In the efficient scenario, the leader is truthful. However, this is not always the case, and protocols need to support intricate fallback strategies, which usually imply node synchronisation and state transfer. 

\looseness=-1 The main disadvantages of BFT protocols, namely that they are slow and expensive to run, support the research of streamlined and cluster-based algorithms. In this sense, researchers developed Hotstuff~\cite{HotStuff}, a streamlined protocol that assumes partial synchrony and uses leader rotation on each block proposal to shift the communication burden from the leader. By using a star-type communication pattern, the protocol achieves linear message complexity and faster response times. Additionally, current research is being conducted to optimise the features of streamlined algorithms~\cite{streamlined-blockchains}, such as Pili~\cite{pili}, Pala~\cite{pala}, Streamlet~\cite{streamlet}, Tendermint~\cite{tendermint} and, previously mentioned, Hotstuff~\cite{HotStuff}. For instance, DAMYSUS improves on top of Hotstuff by reducing the number of communication phases using trusted components, thus achieving better performance~\cite{Damysus}. 

\looseness=-1 Reaching consensus represents a critical point of improvement for distributed protocols. In this sense, the idea of using a weight metric as voting power gained popularity with Proof-of-Stake (PoS) and reputation-based protocols~\cite{repucoin}. Building on top of this kind of mechanism, WHEAT achieved higher performance for state machine replication in geographically distributed settings~\cite{WHEAT}. Next, researchers put together BFT-SMaRt~\cite{BFT-SMART} (an enhanced version of PBFT) and the weighted voting mechanism behind WHEAT to create AWARE, a deterministic, self-monitoring and self-optimising algorithm for reducing the latency of the blockchain~\cite{AWARE}.

So far, research on the benefits of weighted voting has only studied PBFT in AWARE. This paper seeks to address this literature gap by \textit{investigating the impact of weighted voting on streamlined consensus algorithms}. By extending the principles established by AWARE to Hotstuff~\cite{HotStuff} and evaluating the robustness in facing node failures, we aim to contribute to the broader understanding of weighted voting's efficacy in streamlined blockchain systems. 

\looseness=-1 This study consists of a latency prediction model which emulates Hotstuff behaviour to gather data on whether or not applying weighted voting decreases the latency of the blockchain algorithm. By analysing different optimisation techniques, this paper points out possible performance improvements and presents conclusive results that encourage the development of an actual deployment in a further study. 

As an overview, our contributions can be summarised as follows.

\begin{enumerate}
    \looseness=-1 \item We apply AWARE's weighting scheme~\cite{AWARE} to Hotstuff and Chained Hotstuff, using two latency prediction models to estimate the algorithms' performance. 
    
    \item We analyse how optimising the weight distribution to replicas and/or leader rotation impacts latency by employing separate \textit{Simulated Annealing}~\cite{baeldungSimulatedAnnealing} methods.

    \item We explore the possibility of using continuous weight values instead of AWARE's discrete weights. We apply this weighting scheme on Hotstuff whilst ensuring quorum safety and assessing its effectiveness in reducing latency.
\end{enumerate}

\looseness=-1 The rest of this paper is organised as follows.~\cref{related_work} reviews the academic advancements in weighted voting and streamlined algorithms.~\cref{background} presents a thorough description of Hotstuff for the state replication protocol and WHEAT for the underlying weighting scheme, which is also exploited in AWARE.~\cref{methodology} and~\cref{continunous-weighting} describe our contributions by outlining the latency optimisation methods and prediction models.~\cref{experiements_results} delves into the experimental setup and presents our findings.~\cref{responsible_research} reflects on the ethical side of this research.~\cref{future_work} provides a critical overview of our experiments and discusses the next steps that could be taken in this area of research.~\cref{conclusion} concludes this paper with an overview of the impact of studying weighted voting on streamlined blockchain algorithms in the research field.
\section{Related work}\label{related_work}

The literature related to the aforementioned scientific gap revolves around two key concepts: \textbf{weighted voting} and \textbf{streamlined algorithms}. Therefore, we provide an overview of each of the two research areas to gather a better understanding of the improvements that have been achieved over the years.

\textbf{Weighted voting} Inspired by the popularity of Proof-of-Stake protocols, researchers have adapted this idea into using a weight metric as the voting power of nodes in permissioned systems. One of the first projects highlighting the advantages of weighted voting was the Cosmos Network, which used a Tendermint-based blockchain protocol and a stake-based voting approach for reaching consensus~\cite{cosmos}. Next, the possibility of using weights for electing the leader was researched in credit-based PBFT~(CPFT), a blockchain algorithm that tunes the weights based on nodes' past behaviour such that the probability of electing a good leader increases~\cite{CBFT}. Three years later, new research on using vague sets and credit rating for optimising the consensus of credit-based PBFT blockchain algorithms appeared~\cite{cbft-enhanced}. Later on, starting from the same idea of assigning credits to nodes, researchers developed CG-PBFT, a blockchain algorithm that uses a novel credit evaluation model together with a three-way quick sorting algorithm to achieve around $50\%$ increase in throughput~\cite{CG-PBFT}. Moreover, current research has tackled the idea of a reward and punishment system based on node ranking in D-PBFT~\cite{DIANA}.

\textbf{Streamlined algorithms} The research area for streamlined BFT protocols gained interest with the introduction of Hotstuff~\cite{HotStuff}, urging the study of possible optimisations performed on this protocol. Considering multiple points of improvement, researchers came up with variations of Hotstuff targeting a better overall performance of the system. Sync Hotstuff represents one notable research, which shifts the focus from the partially synchronous model of Hotstuff to a fully synchronous one to highlight the trade-offs and impact on performance and security~\cite{synchotstuff}. Next, researchers targeted improving performance by introducing DAMYSUS, a blockchain protocol that enhances Hotstuff by using trusted components to increase resilience and decrease the number of communication phases~\cite{Damysus}. On the same path of decreasing the number of communication steps, Hotstuff-2 represents a two-phase variant which explores the relation between leader responsiveness and liveness for achieving optimal results~\cite{hotstuff-2}. Furthermore, building on top of DAMYSUS, Oneshot is the first streamlined hybrid BFT protocol which achieves the minimal number of communication phases by exploiting the knowledge of the system's state available to the nodes~\cite{oneshot}.

\textbf{Weighted voting on streamlined algorithms} Current research also explored applying the weighted voting scheme showcased in AWARE~\cite{AWARE} to Hotstuff. By leveraging the idea of an adaptive resilience threshold introduced in ThreatAdapt~\cite{ThreatAdapt}, FLASHCONSENSUS~\cite{flash} is optimising AWARE using this mechanism of dynamically reducing the number of replicas that actively participate in the protocol's execution. The research also mentions experiments combining this new algorithm with the weighted voting mechanism. FLASHCONSENSUS-flavoured HotStuff uses weights, best leader selection and smaller quorums to prove the effectiveness of using weighted voting on Hotstuff. However, the research uses the enhanced AWARE algorithm rather than the original, and it also leaves out details regarding the implementation of weighted voting on the streamlined algorithm. In addition, the impact on its chained version and the generalisation from discrete weighting remains unexplored. 

\section{Background} \label{background}
This research aims to combine the state replication protocol of Hotstuff with the weighted scheme introduced by WHEAT~\cite{WHEAT} whilst also considering the optimisation approaches established by AWARE~\cite{AWARE}. Hence, this section delves into the streamlined approach of Hotstuff in~\cref{background_1} and presents the two-weight scheme of WHEAT in~\cref{background_2}.

\subsection{Hotstuff: the streamlined approach} \label{background_1}
Inspired by the simplicity of the theoretical protocol Streamlet~\cite{streamlet} and part of the BFT family, Hotstuff is a blockchain protocol that sets itself apart by achieving linear (in the number of nodes) communication complexity~\cite{HotStuff}. The protocol benefits from the mechanism of switching leaders in each successive round and requires at least $3f + 1$ nodes to tolerate $f$ Byzantine faults.

There are two protocol versions, namely the \textit{Basic Hotstuff} and the \textit{Chained Hotstuff}. The difference between the two comes from the latter's enhanced voting mechanism, which enables a pipelined approach of moving forward multiple blocks in only one round (\textit{view}). For completely processing a block, Hotstuff uses five communication phases, with three core: \textbf{prepare, pre-commit,} and \textbf{commit} and two additional ones: \textbf{new-view}, for sending the last prepared block at the beginning of the protocol, and \textbf{decide}, for actually executing the block at the end (see~\cref{fig:comm phases} in~\cref{appendix1}).

\textbf{New-view} In this first phase, the leader receives the last prepared block and its corresponding view number from each node. This step consists of gathering all the new-view messages triggered at the end of the precedent view.

\textbf{Prepare} This phase concerns finding the next proposal. The leader awaits for $2f + 1$ quorum of nodes with their block-view number information and chooses the block with the highest view number to be extended. Next, the leader sends to all replicas its proposal, and they each provide back a vote if the \textsc{SafeNode} condition is satisfied~\cite{HotStuff}. That is, a replica will accept the proposal if the proposed block extends either its latest locked block or a prepared block from a view higher than the one of its last locked block~\cite{Damysus}. These checks will ensure the safety and liveness of the protocol, respectively.

\textbf{Pre-commit} The proposed block is marked \textit{prepared} as the leader gets $2f + 1$ votes from the replicas, forming a quorum certificate. The leader sends this certificate to all the replicas so that each can verify it, mark the proposed block as prepared and vote for it in the pre-commit phase.

\textbf{Commit} The leader collects again $2f + 1$ votes and forms a quorum certificate of the prepared block, which becomes \textit{locked} at this step. Next, the leader sends this certificate to all replicas to lock the same block, followed by each sending back a vote. 

\textbf{Decide} The leader reaches consensus on the locked block and then \textit{executes} it. All replicas perform the same action after they receive the certificate from the leader.

\textbf{Locking mechanism} The efficiency of communication complexity in the Hotstuff protocol comes with the cost of implementing this locking mechanism. The block is prepared and then locked in the commit phase to preserve the \textit{liveness} of the blockchain consensus algorithm, and only afterwards is it considered \textit{safe} to execute in the final phase.

\subsection{WHEAT: the weighting mechanism} \label{background_2}
Based on the BFT-SMaRt protocol of state machine replication, WHEAT solves the problem of optimising the system's latency in geo-replicated settings~\cite{WHEAT}. The algorithm is able to attain better performance by introducing the concept of additional replicas. In the BFT family, quorums are formed by gathering a majority of replica responses. In contrast, in WHEAT, the size of a quorum is smaller or equal to that of the Byzantine majority.

Effectively, WHEAT uses weighted voting to achieve consensus faster by leveraging the heterogeneity of the wide-area network (WAN). That is, the protocol assigns higher weights to the replicas that yield the lowest end-to-end latency. In this way, they form smaller quorums but account for the majority needed to move forward. The rest of the replicas constitute a fallback strategy since they form a larger quorum that is in place if the faster replicas become idle. Furthermore, AWARE enhances this technology by introducing mechanisms for self-monitoring (deterministic latency prediction) and self-optimisation (voting weights tuning and leader relocation), such that latency is decreased by giving more power to better-performing replicas~\cite{AWARE}.

In the Byzantine Fault Tolerant world, a quorum system represents a collection of subsets of replicas that could possibly form a quorum, and any two subsets intersect by $f + 1$ replicas~\cite{reiter1998byzantine}. By adding $\Delta$ extra replicas and enforcing a quorum formation mechanism relying on weighted replication, WHEAT imposes the subsequent safe weight distribution scheme.

Consider a BFT system of $n$ replicas withstanding a maximum of $f$ faulty and including $\Delta$ additional ones. Hence, $n$ can be expressed as follows:
\begin{equation}
n = 3f + 1 + \Delta \label{eq:number of replicas}
\end{equation}

Furthermore, regarding consensus, each replica should wait for a quorum formation of $Q_v$ weighted votes:
\begin{equation}
Q_v = 2(f + \Delta) + 1 \label{eq:quorum}
\end{equation}

The voting powers take the form of a binary weight distribution over the replicas of WHEAT. Each node has value either $V_{\text{max}}$ or $V_{\text{min}}$, which are computed as follows:
\begin{equation}
V_{\text{max}} = 1 + \frac{\Delta}{f} \label{eq:vmax}
\end{equation}
\begin{equation}
V_{\text{min}} = 1 \label{eq:vmin}
\end{equation}

In the system, the $2f$ replicas that are best performing in terms of latency are attributed weight $V_{\text{max}}$ and all the others take $V_{\text{min}}$. Consequently, the weighted quorum contains at most $n - f$ and at least $2f + 1$ replicas.
\section{Weighted voting for streamlined algorithms} \label{methodology}

To sense the impact of weighted voting on streamlined algorithms, this research funnels on a representative blockchain algorithm, namely Hotstuff~\cite{HotStuff}. This paper evaluates the effectiveness using \textbf{latency decrease} as a measurement metric. Hence, to approach the research problem, a latency prediction model is used to emulate the behaviour of Hotstuff, such that intricate implementation details are ignored, and the focus lies on the predicted time of completing a block proposal. 

Hotstuff employs five communication phases (which together form a view) to execute a block. To predict the latency from the start of the view, when the leader proposes a block, to the end, when the block is executed, the latency prediction model follows the \textbf{new-view, prepare, pre-commit, commit and decide} phases. By analysing a Hotstuff run step-by-step, four quorum formation events can be identified. The leader waits sequentially for \textbf{new-view, prepare, pre-commit and commit} messages to complete a block proposal. Since weighted voting is defined within the consensus mechanism, the latency prediction model concentrates on the quorum formation completion times, which can be optimised. 

Given the required information on the network topology and the weighting scheme (weight assignment for replicas), a latency prediction for a Hotstuff run depends on the latencies registered by the leader. That is, for each message type $x$, the leader retains latency vector $L_x$. In particular, $L_x[i]$ represents the latency, reported by the leader, of receiving the message of type $x$ from replica $i$. Hence, we can use the corresponding latency vectors to compute the times $t_{PREPARE}$, $t_{PRECOMMIT}$, $t_{COMMIT}$, $t_{DECIDE}$ it takes the leader to form a quorum of messages in order to advance to the next phase (see~\cref{Algorithm 1}). By adding together these timestamps, we get the overall predicted latency of completing a Hotstuff view.

Furthermore, to compute the time it takes to form a quorum, the model uses the voting power of a replica expressed by its corresponding weight. In short, when gathering messages for consensus, we first consider the messages that arrive faster, adding up their weights until the quorum formation condition is satisfied. In this way, the latency of the last message that helped constitute the needed quorum equals the overall time it took the leader to reach a consensus, hence advancing to the next phase.

\begin{algorithm}[h]
\small
\SetAlgoNlRelativeSize{-1}
\caption{Latency prediction model for Weighted~Hotstuff}\label{Algorithm 1}
\KwData{weightingScheme, leaderRotation, numberOfViews}
\KwResult{\textbf{total latency} for running in the given network setup}

\vspace{5pt}
latency $\gets$ 0
\vspace{5pt}

\For{viewNumber $\gets$ 0 \KwTo numberOfViews}{
    currentLeader $\gets$ leaderRotation[viewNumber]
    \vspace{5pt}
    
    $L_{new-view}$ $\gets$ getLatencyOfMessages("new-view", currentLeader)
    
    $t_{PREPARE}$ $\gets$ timeToFormQuorum($L_{new-view}$, weights)
    \vspace{5pt}
    
    $L_{prepare}$ $\gets$ getLatencyOfMessages("prepare", currentLeader)
    
    $t_{PRECOMMIT}$ $\gets$ timeToFormQuorum($L_{prepare}$, weights)
    \vspace{5pt}
    
    $L_{precommit}$ $\gets$ getLatencyOfMessages("precommit", currentLeader)
    
    $t_{COMMIT}$ $\gets$ timeToFormQuorum($L_{precommit}$, weights)
    \vspace{5pt}

    $L_{commit}$ $\gets$ getLatencyOfMessages("commit", currentLeader)
    
    $t_{DECIDE}$ $\gets$ timeToFormQuorum($L_{commit}$, weights)
    \vspace{5pt}

    latency $\gets$ latency + ($t_{PREPARE}$ + $t_{PRECOMMIT}$ + $t_{COMMIT}$ + $t_{DECIDE}$)\;
}

\vspace{5pt}
\Return{latency}
\end{algorithm}

\subsection{Weighted Basic Hotstuff} 
\label{weighted-hotstuff}

The basic version of \textbf{Weighted Hotstuff} entails using the $V_{max}$ and $V_{min}$ weights (see Equations~\eqref{eq:vmax} and~\eqref{eq:vmin}) by assigning the highest one to $2f$ replicas. To ensure the safety of the quorum system, the algorithm follows WHEAT~\cite{WHEAT} and introduces the use of $\Delta$, namely the number of additional replicas. In order to get the predicted latency on a given network setting, the weights assigned are fed into~\cref{Algorithm 1}.

\textbf{Best weight assignment} The weight assignment to replicas represents a critical point of improvement. Hence, \textbf{Best Assigned Weighted Hotstuff} employs a \textit{Simulated Annealing}~\cite{baeldungSimulatedAnnealing} approach to find the proper configuration for a network scenario. Starting from the weight distribution used in Weighted Hotstuff, a candidate solution is generated using a perturbation function as follows: find a replica having $V_{min}$ weight and assign it $V_{max}$ instead, whilst its voting power becomes $V_{min}$. This way, multiple weighting schemes are tested by predicting the latency under each possible set of weights as an energy function. Ultimately, the algorithm converges to one weighting scheme, yielding the lowest latency for the network scenario.

\begin{figure}[h]
    \centering
    \includegraphics[width=0.5\textwidth]{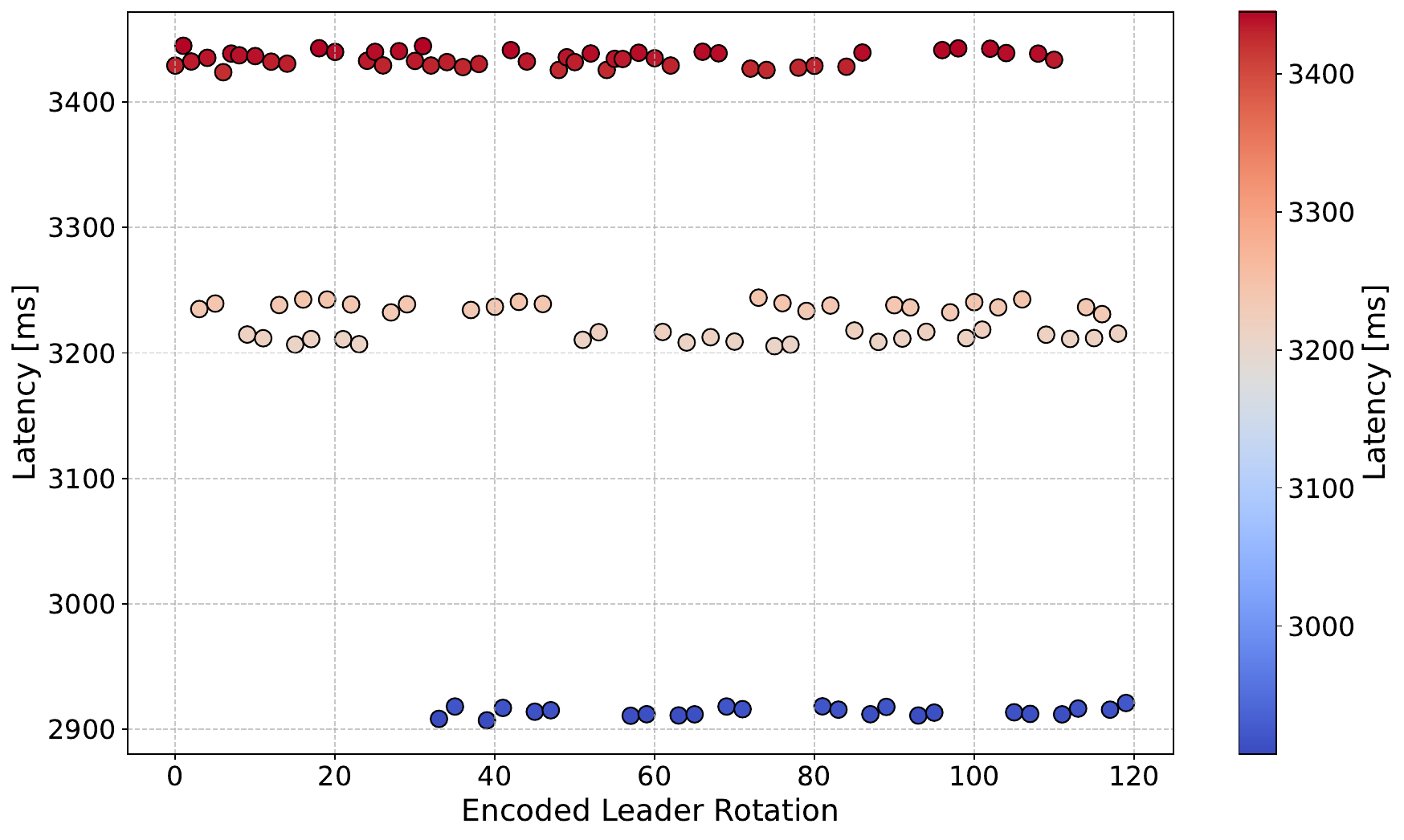}
    \caption{Analysis of the impact of leader rotation on Hotstuff's latency performance for $f = 1, \Delta = 1$, $4$ views executed.}
    \label{fig:analysis_leader_rotation}
\end{figure}

\textbf{Optimal Leader Rotation} The Hotstuff blockchain algorithm benefits from a linear message complexity due to the novel leader rotation scheme it introduces in ledger technologies. Choosing the best possible succession of leaders is a potential improvement point since it strongly correlates with the algorithm's latency (see~\cref{fig:analysis_leader_rotation}). 

However, the leader election itself does not introduce a significant delay directly, but it can indirectly affect the algorithm's behaviour due to leaders' network connectivity, which impacts communication latency. A poorly connected leader can lead to delays in message propagation, and in faultiness cases, it can jeopardise the overall view completion. 

Inspired by AWARE'S established leader relocation mechanism~\cite{AWARE}, this variant of Weighted Hotstuff aims to provide a latency prediction for using an optimised leader rotation. Thus, a decrease in the measurement metric would support further study of a Hotstuff implementation with alternative leader selection strategies. As of its current implementation, Hotstuff uses a \textit{round-robin} leader rotation~\cite{round-robin}. However, a leader selection based on network health, similar to that of AWARE, could be beneficial. Illustrating the optimisation opportunity in Hotstuff, \textit{Simulated Annealing} is used by \textbf{Optimal Leader Rotation Weighted Hotstuff} to run the blockchain protocol against different leader rotation schemes, collecting the best performance that can be achieved in a given network setting. 

The metaheuristic method starts from the classical leader rotation of Hotstuff, which serves as a baseline for comparing alternative solutions. From the existing state, the algorithm moves to a neighbouring one by swapping two leader positions, generating a new possible leader rotation scheme. If the new one proves to be better, the simulation moves to this new state and continues in the same manner until it converges to a leader sequence that is most suitable in terms of latency for a proposed scenario.

\textbf{Optimal Leader Rotation + Best Assigned} A combined \textit{Simulated Annealing} approach targeting both optimisation strategies illustrates their latency reduction impact on Weighted Hotstuff. The algorithm generates a candidate that follows with probability $\frac{1}{2}$ the weighting distribution optimisation model and, with the same probability, a leader rotation one. Hence, the solution indicates the best weighting scheme and leader rotation based on the network environment.

\subsection{Weighted Chained Hotstuff} \label{chained-weighted-hotstuff}

The Chained Hotstuff blockchain algorithm represents a pipelined version of the original one. It benefits from the similarity of the five communication phases to introduce a unique approach to advancing to the next phase on multiple blocks. In a network with $n$ replicas, each one votes for the progress of at most $n$ block proposals in one view. Hence, a leader proposes a block in view $i$ and forwards the responsibility to the leader of the upcoming view until, ideally, it is executed in view $i + 4$ by the corresponding leader.

The latency prediction model used in Weighted Hotstuff is tweaked accordingly to account for these changes in state replication behaviour. For one view, we have a set of block proposals $b_1, b_2...b_n$ for which we consider simultaneous advancement. Since the focus still lies on quorum formation times, the $L_{x,b_j}[i]$ vector encapsulates the latency reported by the leader of the view for receiving the message of type $x$ for block proposal $b_j$ from replica $i$. The latency vectors are then used to predict the time it takes to reach the required quorum weight for consensus. Note that due to its state replication nature, the adapted model exhibits a warming-up period of $n$ views until the simulation reaches the point of having a maximum capacity of block proposals in each upcoming view.

As in the Weighted Hotstuff algorithm, its chained version also employs a weight metric as voting power in the consensus mechanism. The \textbf{Weighted Chained Hotstuff} assigns $V_{max}$ and $V_{min}$ weights to $2f$ replicas in the network. Furthermore, the blockchain algorithm benefits from the same possible performance improvements: weighting distribution and leader rotation. In this sense, this research explores their impact by developing two \textit{Simulated Annealing} approaches: \textbf{Best Assigned Weighted Chained Hotstuff} and \textbf{Optimal Leader Rotation Weighted Chained Hotstuff}, which follow the same paradigm presented in~\cref{weighted-hotstuff}, but with the required changes for emulating the proper chaining behaviour of the protocol.

\section{Continuous Weighting Scheme} \label{continunous-weighting}

WHEAT~\cite{WHEAT} presents the effectiveness of using two-weight vote power assignments in optimising blockchain systems' performance in geographically distributed environments. The use of $V_{max}$ and $V_{min}$ can be interpreted as a limitation of the algorithm. Following this idea, \textbf{Continuous Weighted Hotstuff} extends on top of the Weighted Hotstuff algorithm by introducing a weighting scheme generalisation using \textit{Simulated Annealing}. This research only tests its efficacy on Hotstuff to compare its performance to the other variants of protocol optimisations. Nevertheless, the continuous weighting scheme is not limited to the streamlined blockchain world and could be applied to any voting power-based consensus algorithm. 

The \textit{Simulated Annealing} approach for finding the continuous weighting scheme that achieves minimal latency works as follows. The latency prediction model of~\cref{Algorithm 1} is used as an energy function to evaluate the fitness of alternative weighting schemes. To generate a new candidate from the current state, the set of weights uses a $perturbationStep = 0.1$ hyperparameter to draw a new weight from the uniform distribution $U(currentWeight - perturbationStep, currentWeight + perturbationStep)$. Moreover, the voting power of replicas is capped between zero and two to reduce the searching area of the annealing process. The required sum of weights equals the weighted quorum condition of a consensus process to progress on the block proposal safely. Hence, the weighted quorum size depends on the replicas' specific weighting distribution scheme. The choice of capping the weight values does not represent a limitation of the research conducted and does not endanger the validity of the optimisation since the weights are relative to the network scenario.

For a quorum to be \textit{safe}, the following properties of the quorum system need to be validated:

\begin{enumerate}
    \item \textbf{Availability:} Even when the most powerful $f$ replicas fail, there is at least one quorum to reach consensus.
    \item \textbf{Consistency:} Every two quorums overlap by at least one correct replica.
\end{enumerate}

The probabilistic model uses an additional functionality to deem the correctness of the two quorum system conditions when predicting latency. When a new candidate is used to predict the latency, its corresponding quorum size must be computed. First, considering the $f$ replicas having the highest voting power faulty implies that the sum of weights of the remaining replicas is an upper bound on the quorum size. \textit{Availability} is satisfied by considering this upper bound as the required weighted consensus condition. To ensure \textit{consistency}, only the subsets of replicas with total voting power greater or equal to the weighted quorum size are valid. The algorithm takes any two valid quorums and verifies if they overlap by $f + 1$ replicas. Suppose the candidate continuous weighting scheme passes all of these steps. In that case, the quorum system is well-founded, and the latency predicted by the model is taken into account by the annealing process in moving to a subsequent state. Otherwise, the current state remains unmodified for the next step of the algorithm.

\section{Evaluation} \label{experiements_results}

This paper analyses the impact of weighted voting on streamlined algorithms by gathering data from multiple experiments to determine whether or not introducing the voting power assignment generates significant latency improvements. 

\subsection{Experimental setup}

We evaluate the performance of Weighted Hotstuff~(\cref{weighted-hotstuff}), Weighted Chained Hotstuff~(\cref{chained-weighted-hotstuff}) and their multiple optimisation variants to weigh up against those of basic and chained Hotstuff. Thus, we perform experiments in two scenarios: non-faulty, when the network behaves normally, and faulty, when some replicas are idle. 

To run the latency prediction models, we use Python scripts to test the behaviour of protocols in a given network scenario. Hence, each experiment depends on the specified distributed environment. Moreover, the latency prediction algorithm for Weighted Hotstuff requires providing a set of weights. The algorithm's behaviour matches that of Best Assigned Weighted Hotstuff in case of optimal weight assignment. Thus, the experiments are conducted by tailoring the weighting distribution to the network scenario. That is, data on the distance within all replicas is gathered, and the $f$ best and $f$ worst connected replicas are assigned $V_{max}$ voting power. 

To make the analysis reliable, experiments are conducted using real data collected from cloudping, a tool for latency monitoring of deployed AWS clusters~\cite{cloudping}. For the results presented later in this section, the following clusters were used: \textit{Cape Town (af-south-1), Hong Kong (ap-east-1), Canada (ca-central-1), London (eu-west-2) and Northern California (us-west-1)}. This choice of network setting supports the research of weighted voting efficacy on geographically distributed settings since it reflects the heterogeneity of wide area network (WAN) environments. Moreover, simulations were performed using a blockchain system with $f = 1$ and $\Delta = 1$ (having a total of $n = 5$ replicas). Since Hotstuff is a protocol that sets itself apart from the others in the BFT family by using a new leader in each view, the views represent a critical aspect of the influence of weighted voting. Thus, we vary the number of views by performing simulations for all values from $5$ to $20$, focusing on the average latency per view.

The latency prediction models depend on the values generated for latency vectors $L_x$. In an actual deployment, the latency at which the leader receives messages from each replica is influenced by two factors: distance from leader to replica, reflected in link latency and the payload of the transmitted data, namely message delay. To emulate the network behaviour properly, for a leader $i$, we create its corresponding latency vectors $L_x$ as follows: distance within clusters (for the link latency) plus an offset drawn from a uniform sample $U(0, 5)$ for $L_{new-view}$ (since new-view messages encapsulate the block proposal, whereas the rest have lower payload containing just hashes) and $U(0, 2)$ for the rest. If the link is fast enough, the message payload latency can be neglected. However, for a better analysis of weighted voting, we went for a uniform distribution with significantly lower values than the cluster latency. In this way, message complexity is considered without severely influencing the protocol's behaviour but with the perks of mocking the heterogeneity of real environments.

\subsection{Non-faulty scenario}
The experiments performed considered as baseline Basic and Chained Hotstuff whose latencies are obtained by running the prediction model of Weighted Hotstuff with all weights equal to $V_{normal} = 1$. In this way, we highlight the effectiveness of weighted voting on streamlined protocols (see~\cref{tbl:latency_improvement_table}). 

\begin{table}[h]
\centering
\caption{Performance comparison between Hotstuff and Weighted Hotstuff variants based on \cref{fig:hotstuff,fig:chained hotstuff}.}
\label{tbl:latency_improvement_table}
\small
\begin{tabular}{@{}lcc@{}}
\toprule
 & \textbf{Basic} & \textbf{Chained} \\
\midrule
\textbf{No optimisation}              & -7.13\%                    & -7.28\%    \\          
\textbf{Best Assigned}                  & -19.48\%                   & -19.64\% \\
\textbf{Optimal Leader Rotation}         & -10.52\%                   & -9.78\% \\           \textbf{Continuous}            & -18.30\%                   & n.a     \\          \textbf{Optimal Leader Rotation + Best Assigned}  & -24.36\%                   & -22.48\%   \\       
\bottomrule
  \end{tabular}
\end{table}

\begin{figure}[h]
    \includegraphics[width=0.5\textwidth]{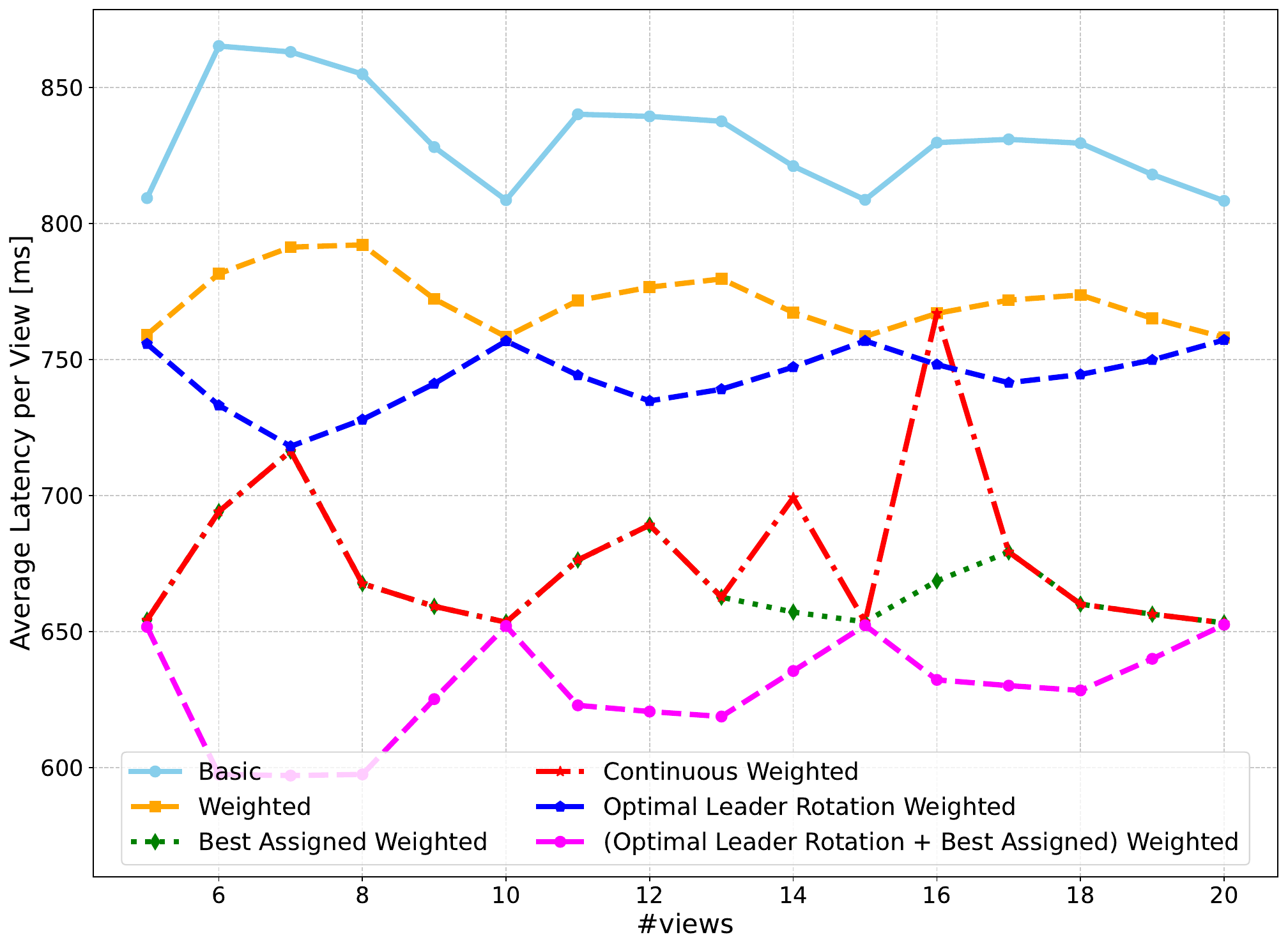}
    \caption{Average latency per view in Hotstuff protocol variants for $f = 1, \Delta = 1$.}
    \label{fig:hotstuff}
\end{figure}

\textbf{Weighted Hotstuff}~\cref{fig:hotstuff} showcases the impact of weighted voting on Hotstuff. It is clear from this visual representation of latency performance that introducing weighted voting decreases the latency of running the streamlined blockchain algorithm. Compared with the Basic Hotstuff in which all weights are assigned equal voting power, Weighted Hotstuff exhibits around $7\%$ decrease in latency. Adding to the $V_{max}/V_{min}$ weight assignment, the Optimal Leader Rotation variant reduces latency by up to~$10\%$. However, there is room for improvement as Best Assigned and Continuous Weighted Hotstuff both indicate almost $20\%$ latency decrease. Furthermore, by combining the two optimisation approaches, (Optimal Leader Rotation + Best Assigned) Weighted Hotstuff reveals around $25\%$ latency reduction. Hence, the prediction models developed on multiple protocol variants support the idea established in AWARE of introducing leader relocation and weight-tuning mechanisms for improving the performance of blockchain algorithms. 

~\cref{fig:hotstuff} also shows that the optimal leader rotation variant has the same latency as its underlying weight optimisation Hotstuff variant over multiple of $n$ views. That is a consequence of the simulation framework using a static network setting for the experiments conducted. Since for such a simulation consisting of a multiple of $n$ views, each replica is the leader for an equal amount of times, the cumulative latency over all the views is the same.

\begin{figure}[h]
    \centering
    \includegraphics[width=0.5\textwidth]{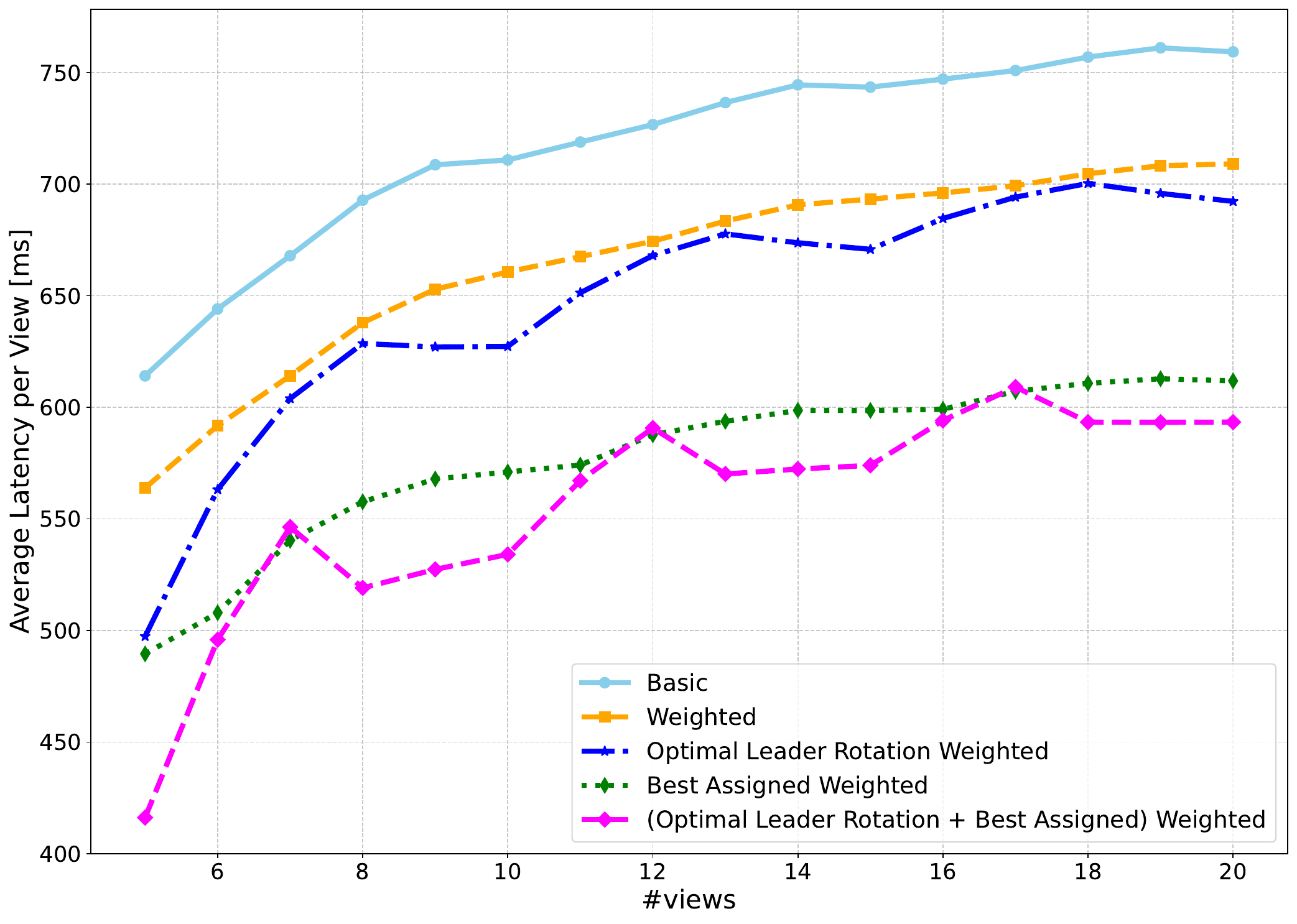}
    \caption{Average latency per view in Chained Hotstuff protocol variants for $f = 1, \Delta = 1$.}
    \label{fig:chained hotstuff}
\end{figure}

\textbf{Chained Weighted Hotstuff}~\cref{fig:chained hotstuff} showcases the impact of weighted voting on Chained Hotstuff. The figure highlights the warming-up period of $n$ views and supports the research of weighted voting on streamlined algorithms with a $7\%$ latency improvement gained by just assigning weights. Furthermore, adding the leader rotation optimisation on top of it accounts for an additional $3\%$ decrease in latency. As presented in Best Assigned Weighted Hotstuff, its chained version follows the same trends, exhibiting an average latency per view with $20\%$ better than the baseline and performing best with an up to $25\%$ reduction for the combined version of the two optimisations.

\subsection{Faulty scenario}
Resilience against failures is critical in any distributed environment.  Hence, we compare the performance of different Weighted Hotstuff optimisations under faulty conditions. Since streamlined protocols are part of the BFT family, the system would need at least $3f + 1 $ nodes to withstand $f$ failures. Thus, for a given network scenario, the $f$ replicas holding the highest voting power are considered idle, jeopardising faster consensus. In this way, we force the analysed streamlined algorithms to mimic their fallback scenario strategy.

\begin{figure}[h]
    \includegraphics[width=0.5\textwidth]{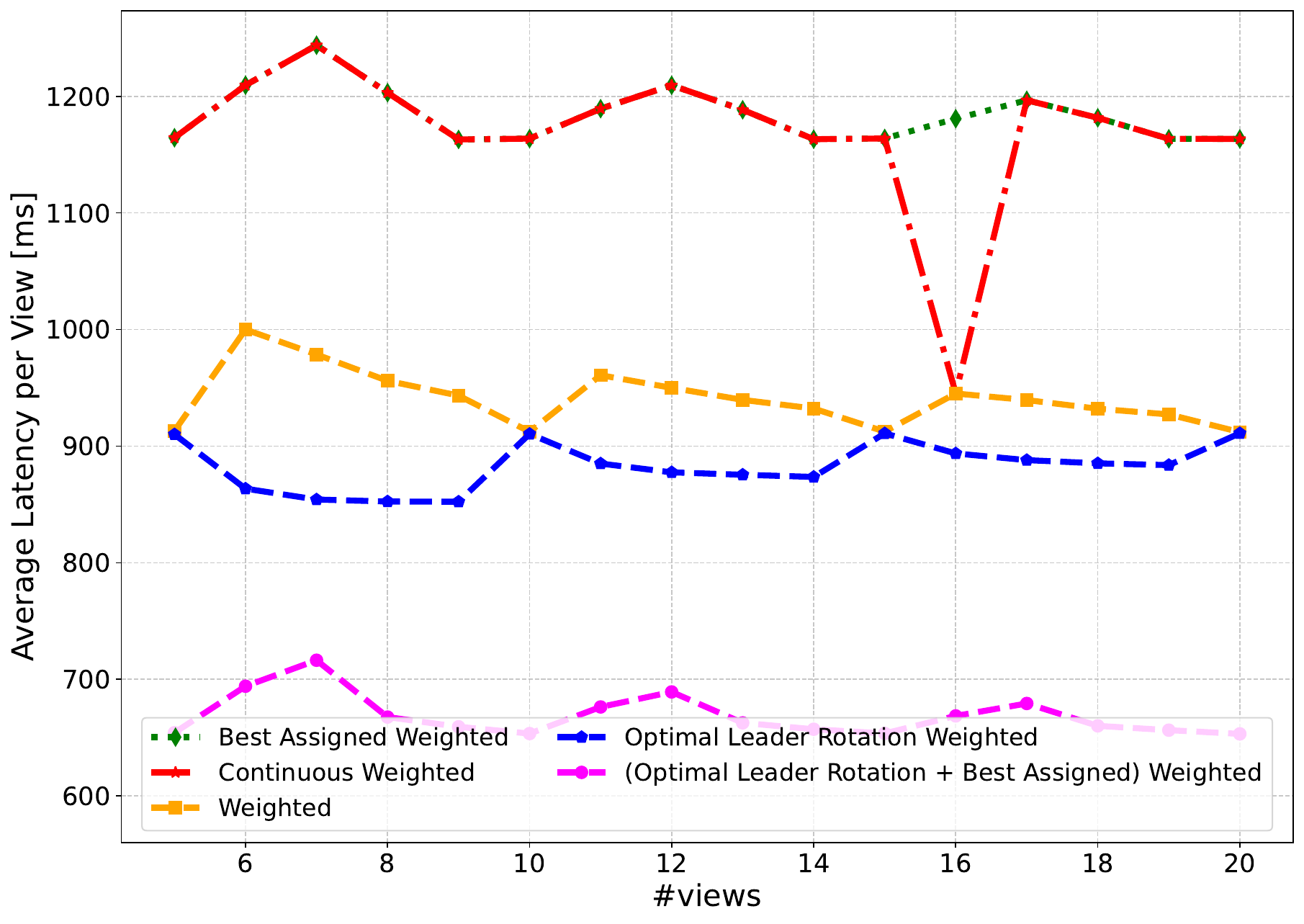}
    \caption{Average latency per view in Hotstuff protocol variants for \textit{faulty scenario}, $f = 1, \Delta = 1$.}
    \label{fig:hotstuff faulty}
\end{figure}

\textbf{Weighted Hotstuff}~\cref{fig:hotstuff faulty} shows that, out of all, the (Optimal Leader Rotation + Best Assigned) Weighted Hotstuff performs the best. Conversely, Best Assigned and Continuous Weighted Hotstuff express their over-fitting nature by showcasing significantly higher fallback latency. Since these two variants optimise the weight assignment for a given network scenario, the system would inevitably take longer to recover from an idle actor. Additionally, they perform worse than Weighted Hotstuff and its leader rotation optimisation version under normal network conditions and far worse under faulty ones.

\begin{figure}[h]
    \centering
    \includegraphics[width=0.4\textwidth]{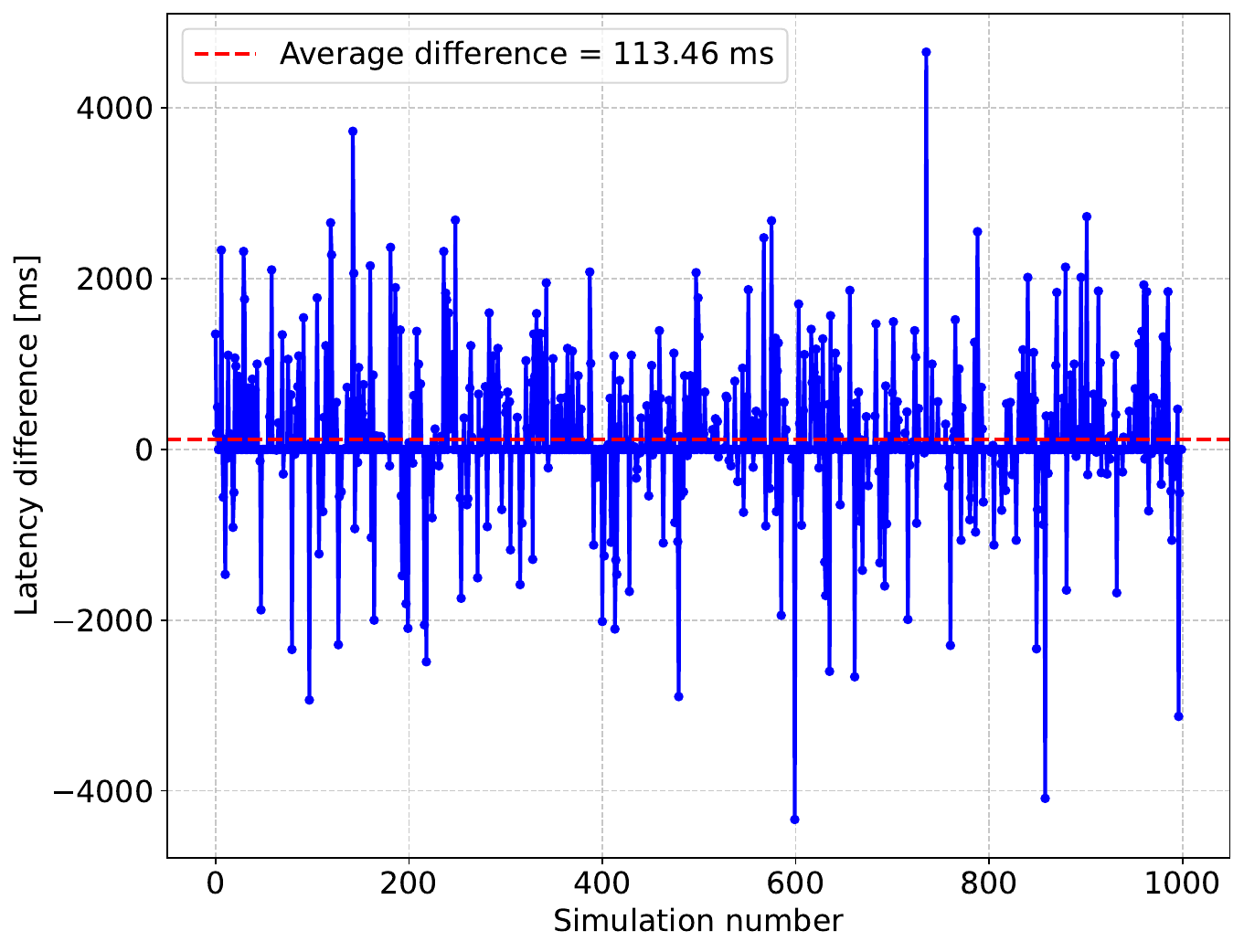}
    \caption{Difference in latency performance between Best Assigned and Continuous Weighted Hotstuff variants for 1000 \textit{faulty scenario} simulations, $f = 1, \Delta = 1$, $10$ views executed.}
    \label{fig:analysis_continuous_hotstuff}
\end{figure}

\textbf{Continuous Weighting Scheme} As expected, Continuous Weighted Hotstuff performs at least as well as the Best Assigned one in non-faulty environments. However, we introduce this generalisation to investigate its impact in faulty scenarios. For this, we experimented with multiple simulations on randomly generated network topologies with within clusters latency ranging between $0$ and \SI{400}{\milli\second}.~\cref{fig:analysis_continuous_hotstuff} showcases the difference in latency between the Best Assigned and Continuous Weighted Hotstuff and proves that the latter performs better or equal in $85\%$ of the performed simulations (see~\cref{fig:continuous comparison percentages} in~\cref{appendix2}). Notably, the continuous variant fails to surpass the discrete one in all simulations due to the \textit{Simulated Annealing} algorithm terminating before reaching the global minimum. Hence, this research supports further study of the continuous metric for voting power assignment. 

\begin{figure}[h]
    \centering
    \includegraphics[width=0.5\textwidth]{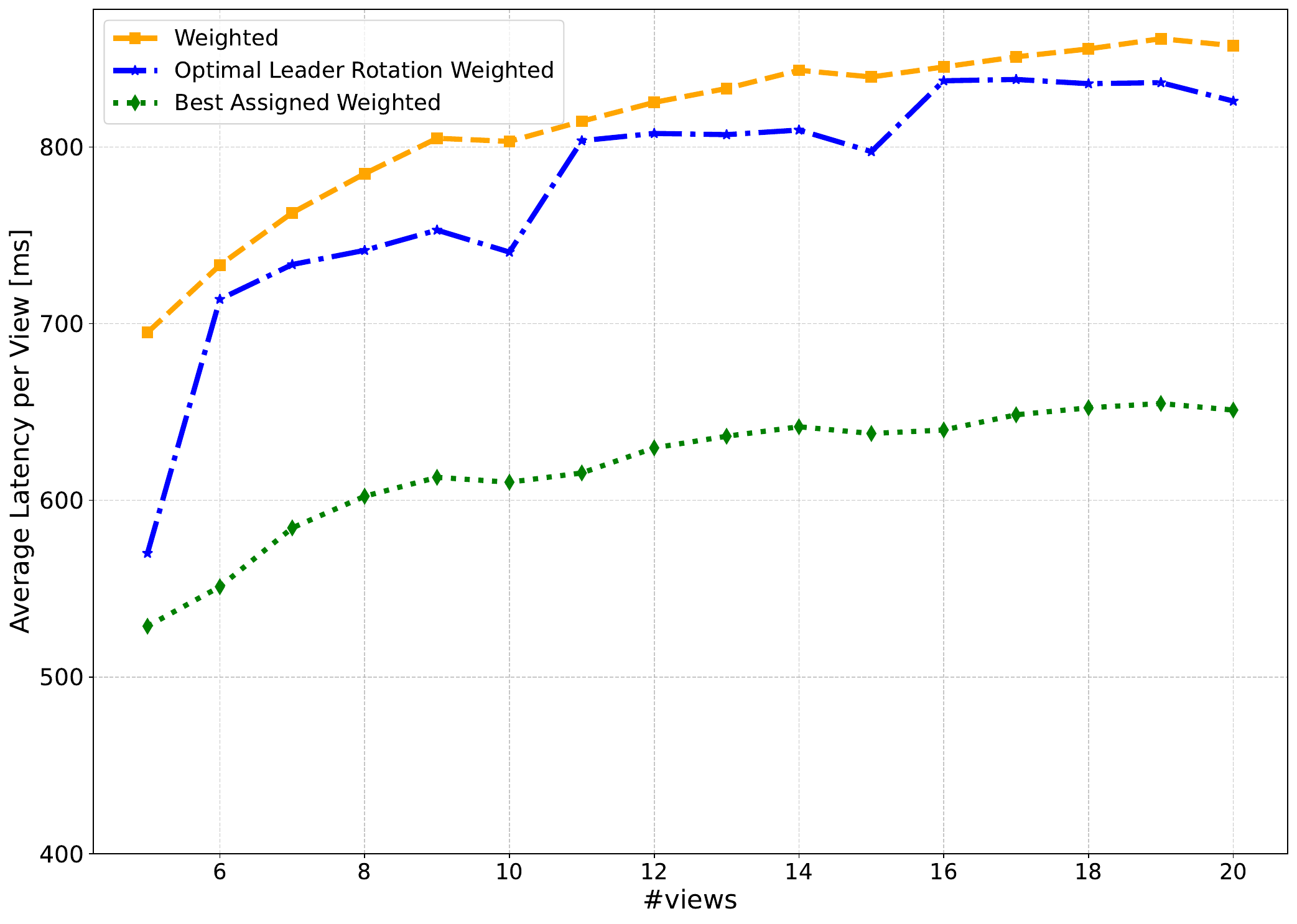}
    \caption{Average latency per view in Chained Hotstuff protocol variants for \textit{faulty scenario}, $f = 1, \Delta = 1$.}
    \label{fig:chained hotstuff faulty}
\end{figure}

\textbf{Chained Weighted Hotstuff} Despite behaving similarly to its basic version, Chained Weighted Hotstuff and its optimisations show major differences in fallback behaviour, as presented in~\cref{fig:chained hotstuff faulty}. Best Assigned Weighted Chained Hotstuff has the lowest fallback latency. That is due to the communication pattern of the chained protocol. Because simulations are performed under a specific set of weights, Chained Hotstuff cannot overfit when optimising for best weight assignment since the pipelined block proposal mechanism heavily influences the algorithm performance. With voting quorums for multiple blocks in one view, some weight assignments might benefit the consensus of one block whilst requiring more time to gather messages for the other. Hence, the Best Assigned Chained Hotstuff has higher resilience against idle nodes. Furthermore, the Weighted and Optimal Leader Rotation Weighted Hotstuff have higher fallback latency but still within the expected limits of \SI{200}{\milli\second} delay, with the latter performing better under this faulty environment setting.

\section{Responsible Research} \label{responsible_research}

This section reflects upon the ethical aspects of the research, presenting the measures taken to adhere to the Netherlands Code of Conduct for Research Integrity~\cite{codeofconduct}. 

\textbf{Datasets} Given the area of research, namely blockchain algorithms, this paper does not use predefined datasets or extend upon implemented algorithms of previous academic work. This research employs designing latency prediction models for Weighted Hotstuff and Weighted Chained Hotstuff alongside their variants. As detailed in~\cref{experiements_results}, we used data from cloudping~\cite{cloudping} to get the latency between different AWS clusters. Moreover, we motivated our choice of clusters, providing transparency in our research. Hence, the experiments aim to emulate a real-life scenario using reliable data to support the study of weighted voting on streamlined algorithms.

\textbf{Results} Due to the nature of this research project, results are heavily influenced by the prediction models. Hence, we provide a Gitlab repository that describes the codebase extensively in its included README file~\cite{gitlab-repo}. Furthermore, all design choices are mentioned and explained in~\cref{methodology} and~\cref{experiements_results}. Besides, the results can be obtained by running the Python experiment files. In interpreting the results, we objectively point out the performance aspects presented in the figures and explain any odd behaviour that the plots showcase. This way, we target transparency and reproducibility of our research, complying with the FAIR (Findable, Accessible, Interoperable and Reusable) principles.

\textbf{Research process} This research process was conducted responsibly by ensuring a proper literature review.~\cref{related_work} describes all the research performed in recent years on our study's two areas of interest: weighted voting and streamlined algorithms. Additionally, we mentioned a research study that briefly looked into the possibility of applying weighted voting on Hotstuff, and we discussed its limitations to put our research endeavours into perspective. The latency prediction models were developed after an in-depth analysis of Hotstuff and Chained Hotstuff communication protocols to ensure the validity of our solutions. Moreover, for Continuous Weighted Hotstuff, we followed the axiomatic quorum system properties and provided an implementation that would comply with the availability and consistency requirements. Throughout the research process, we tried to be as meticulous as possible to guarantee the ethics and correctness of our paper.

\section{Discussion} \label{future_work}

The results presented in~\cref{experiements_results} showcase the impact of our research on the study of streamlined blockchain algorithms. Providing multiple prediction models for different optimisations of weighted voting in Hotstuff and Chained Hotstuff, this paper states the efficacy of the voting power mechanism in such ledger technologies. Thus, we provide real arguments towards the broader study of this research area.

\textbf{Limitations} Some experiments, such as simulations of the \textit{Simulated Annealing} approaches for $n > 15$ replicas, are impractical due to the high code complexity. Moreover, for the continuous weighting scheme, the function for checking that the properties of the quorum are satisfied is computationally expensive, making simulations for $n > 4$ infeasible~(see~\cref{fig:continuous times} in~\cref{appendix2}). Hence, further research could benefit from extensive analysis for latency prediction in a broader network context.

Following the same limitation pattern, the latency prediction models only treat the scenario of running protocols on a specific set of weights. The next step would be dynamically changing the weights from one view to the other to accommodate the leader rotation better. However, to ensure the availability and consistency of the quorum system, this approach entails checking that every two quorums formed on a protocol run overlap by at least one correct replica. This check only would severely delay the overall simulation time of the experiment. Thus, a study should be conducted to circumvent this limitation and research a safe way of changing the voting powers.

Furthermore, the experiments benefit from using specified reliable data of within AWS clusters latency extracted from cloudping~\cite{cloudping}. Even though the setting supports this research, it also introduces some limitations. The most notable one is for the study of the continuous weighting scheme approach, which points out the significant impact of the technique only when run on a randomly generated network scenario. Hence, the next step would be extending this research by running the prediction models on timestamped data from AWS cluster monitoring to observe how weighted voting would impact performance given real-time network fluctuations.

\textbf{Future work} This paper aims to provide the means to understand the impact of weighted voting on streamlined blockchain algorithms. Hence, we advocate for future studies in this area by showing results of possible latency improvements generated through the weighted voting approach. In this sense, upcoming research should apply weighted voting on the actual implementation of Hotstuff and Chained Hotstuff protocols. Instead of using the cloudping data, the research should deploy AWS clusters to observe the real-time behaviour of the blockchain algorithms. Furthermore, the research should follow AWARE's steps of establishing leader relocation and vote assignment fine-tuning mechanisms for improving latency performance in geographically distributed settings whilst taking into account the unreliable side of distributed environments.

\section{Conclusion} 
\label{conclusion}

This paper explored the potential of weighted voting in optimising streamlined blockchain algorithms and analysed, in particular, the latency impact on Hotstuff and its chained variant~\cite{HotStuff}. It investigated possible protocol improvements in weight assignment and leader rotation. Besides, it looked into extending from AWARE's weighting scheme~\cite{AWARE} to a generalised continuous approach that would improve performance whilst maintaining the safety of the quorum system. To this end, we introduced latency prediction models that emulate Hotstuff and Chained Hotstuff behaviour and assign voting powers to the system's nodes, hence simulating a \textbf{Weighted (Chained) Hotstuff} protocol. Furthermore, we used \textit{Simulated Annealing} to estimate the impact of optimising the weight assignment distribution in \textbf{Best Assigned Weighted} and the leader rotation in \textbf{Optimal Leader Rotation Weighted} for both basic and chained versions of the streamlined protocol. Using these latency prediction models, we conducted experiments on reliable data of within AWS clusters latency from cloudping~\cite{cloudping} and also analysed the faulty scenario of best-performing nodes becoming idle. In this way, our experiments are based on real data and consider the fallback scenario, producing results which support future research in the area. Hence, we showcased that only applying weighted voting to a streamlined blockchain algorithm reduces latency by around $7\%$. Moreover, combining best weight assignment and optimal leader rotation achieves minimal latency, almost $25\%$ lower than one of the classic protocols. As for the \textbf{Continuous Weighted Hotstuff}, the enhancement from the discrete set of weights performs in faulty settings equally well or better than Best Assigned Weighted Hotstuff in around $85\%$ of the simulations.

In short, this research represents the standing proof that weighted voting decreases the latency of Hotstuff and Chained Hotstuff, and optimisations of the weight distribution and leader rotation are critical for further performance improvement. As for the generalisation of AWARE's weighting scheme~\cite{AWARE}, this paper introduces a novel approach of using continuous weights as the voting power of the nodes, which is set to improve performance in recovery scenarios. Even though applied in this study only on the Hotstuff algorithm, any blockchain algorithm can employ the continuous weighting scheme. The results provided in this research, together with the novel ideas described, are a founding base for the study of weighted voting in streamlined algorithms and its shift from the discrete model.

\bibliographystyle{IEEEtran}
\bibliography{main}

\begin{thebibliography}{10}
\providecommand{\url}[1]{#1}
\csname url@samestyle\endcsname
\providecommand{\newblock}{\relax}
\providecommand{\bibinfo}[2]{#2}
\providecommand{\BIBentrySTDinterwordspacing}{\spaceskip=0pt\relax}
\providecommand{\BIBentryALTinterwordstretchfactor}{4}
\providecommand{\BIBentryALTinterwordspacing}{\spaceskip=\fontdimen2\font plus
\BIBentryALTinterwordstretchfactor\fontdimen3\font minus \fontdimen4\font\relax}
\providecommand{\BIBforeignlanguage}[2]{{%
\expandafter\ifx\csname l@#1\endcsname\relax
\typeout{** WARNING: IEEEtran.bst: No hyphenation pattern has been}%
\typeout{** loaded for the language `#1'. Using the pattern for}%
\typeout{** the default language instead.}%
\else
\language=\csname l@#1\endcsname
\fi
#2}}
\providecommand{\BIBdecl}{\relax}
\BIBdecl

\bibitem{SMR}
F.~B. Schneider, ``{Implementing fault-tolerant services using the state machine approach: A tutorial},'' \emph{ACM Computing Surveys (CSUR)}, vol.~22, no.~4, pp. 299--319, 1990.

\bibitem{engcons}
{R. Asif and S. R. Hassan}, ``{Shaping the future of Ethereum: exploring energy consumption in Proof-of-Work and Proof-of-Stake consensus},'' \emph{{Frontiers in Blockchain}}, vol.~6, 2023.

\bibitem{ThreatAdapt}
D.~S. Silva, R.~Graczyk, J.~Decouchant, M.~V{\"o}lp, and P.~Esteves-Verissimo, ``{Threat adaptive byzantine fault tolerant state-machine replication},'' pp. 78--87, 2021.

\bibitem{amir2010prime}
Y.~Amir, B.~Coan, J.~Kirsch, and J.~Lane, ``Prime: Byzantine replication under attack,'' \emph{IEEE transactions on dependable and secure computing}, vol.~8, no.~4, pp. 564--577, 2010.

\bibitem{PBFT}
M.~Castro, B.~Liskov \emph{et~al.}, ``{Practical byzantine fault tolerance},'' in \emph{OsDI}, vol.~99, no. 1999, 1999, pp. 173--186.

\bibitem{lamport2019byzantine}
L.~Lamport, R.~Shostak, and M.~Pease, ``{The Byzantine generals problem},'' in \emph{{Concurrency: The Works of Leslie Lamport}}, 2019, pp. 203--226.

\bibitem{cachin2017blockchain}
C.~Cachin and M.~Vukolic, ``{Blockchain Consensus Protocols in the Wild},'' \emph{CoRR}, vol. abs/1707.01873, 2017.

\bibitem{natoli2019deconstructing}
C.~Natoli, J.~Yu, V.~Gramoli, and P.~J.~E. Ver{\'{\i}}ssimo, ``{Deconstructing Blockchains: {A} Comprehensive Survey on Consensus, Membership and Structure},'' \emph{CoRR}, vol. abs/1908.08316, 2019.

\bibitem{HotStuff}
M.~Yin, D.~Malkhi, M.~K. Reiter, G.~G. Gueta, and I.~Abraham, ``{HotStuff: BFT consensus with linearity and responsiveness},'' in \emph{Proceedings of the 2019 ACM Symposium on Principles of Distributed Computing}, 2019, pp. 347--356.

\bibitem{streamlined-blockchains}
E.~Shi, ``{Streamlined blockchains: A simple and elegant approach (a tutorial and survey)},'' in \emph{Advances in Cryptology--ASIACRYPT 2019: 25th International Conference on the Theory and Application of Cryptology and Information Security, Kobe, Japan, December 8--12, 2019, Proceedings, Part I 25}.\hskip 1em plus 0.5em minus 0.4em\relax Springer, 2019, pp. 3--17.

\bibitem{pili}
T.~H. Chan, R.~Pass, and E.~Shi, ``{Pili: An extremely simple synchronous blockchain},'' \emph{Cryptology ePrint Archive}, 2018.

\bibitem{pala}
------, ``{Pala: A simple partially synchronous blockchain},'' \emph{Cryptology ePrint Archive}, 2018.

\bibitem{streamlet}
B.~Y. Chan and E.~Shi, ``{Streamlet: Textbook streamlined blockchains},'' in \emph{Proceedings of the 2nd ACM Conference on Advances in Financial Technologies}, 2020, pp. 1--11.

\bibitem{tendermint}
E.~Buchman, J.~Kwon, and Z.~Milosevic, ``{The latest gossip on {BFT} consensus},'' \emph{CoRR}, vol. abs/1807.04938, 2018.

\bibitem{Damysus}
J.~Decouchant, D.~Kozhaya, V.~Rahli, and J.~Yu, ``{DAMYSUS: streamlined BFT consensus leveraging trusted components},'' in \emph{Proceedings of the Seventeenth European Conference on Computer Systems}, 2022, pp. 1--16.

\bibitem{repucoin}
J.~Yu, D.~Kozhaya, J.~Decouchant, and P.~Esteves-Verissimo, ``{Repucoin: Your reputation is your power},'' \emph{IEEE Transactions on Computers}, vol.~68, no.~8, pp. 1225--1237, 2019.

\bibitem{WHEAT}
J.~Sousa and A.~Bessani, ``{Separating the WHEAT from the chaff: An empirical design for geo-replicated state machines},'' in \emph{2015 IEEE 34th Symposium on Reliable Distributed Systems (SRDS)}.\hskip 1em plus 0.5em minus 0.4em\relax IEEE, 2015, pp. 146--155.

\bibitem{BFT-SMART}
A.~Bessani, J.~Sousa, and E.~E. Alchieri, ``{State machine replication for the masses with BFT-SMART},'' in \emph{2014 44th Annual IEEE/IFIP International Conference on Dependable Systems and Networks}.\hskip 1em plus 0.5em minus 0.4em\relax IEEE, 2014, pp. 355--362.

\bibitem{AWARE}
C.~Berger, H.~P. Reiser, J.~Sousa, and A.~Bessani, ``{AWARE: Adaptive wide-area replication for fast and resilient Byzantine consensus},'' \emph{IEEE Transactions on Dependable and Secure Computing}, vol.~19, no.~3, pp. 1605--1620, 2020.

\bibitem{baeldungSimulatedAnnealing}
\BIBentryALTinterwordspacing
Baeldung, ``{Simulated Annealing},'' Website, 2023, {Accessed on May 30, 2024}. [Online]. Available: \url{https://www.baeldung.com/cs/simulated-annealing}
\BIBentrySTDinterwordspacing

\bibitem{cosmos}
\BIBentryALTinterwordspacing
J.~Kwon and E.~Buchman, ``{The Cosmos Network: A Primer on Tendermint-Based Blockchains},'' Tendermint Inc. and Interchain GmbH, White Paper, 2016. [Online]. Available: \url{https://github.com/cosmos/cosmos/blob/master/WHITEPAPER.md}
\BIBentrySTDinterwordspacing

\bibitem{CBFT}
Y.~Wang, Z.~Song, and T.~Cheng, ``{Improvement research of PBFT consensus algorithm based on credit},'' in \emph{Blockchain and Trustworthy Systems: First International Conference, BlockSys 2019, Guangzhou, China, December 7--8, 2019, Proceedings 1}.\hskip 1em plus 0.5em minus 0.4em\relax Springer, 2020, pp. 47--59.

\bibitem{cbft-enhanced}
Z.~Zeng, B.~Wen, W.~Du, F.~Zhang, and W.~Zhou, ``{PBFT Consensus Algorithm Optimization Scheme Based on Vague Sets and Credit Rating},'' in \emph{2023 6th International Conference on Software Engineering and Computer Science (CSECS)}.\hskip 1em plus 0.5em minus 0.4em\relax IEEE, 2023, pp. 1--5.

\bibitem{CG-PBFT}
J.~Liu, X.~Deng, W.~Li, and K.~Li, ``{CG-PBFT: an efficient PBFT algorithm based on credit grouping},'' \emph{Journal of Cloud Computing}, vol.~13, no.~1, pp. 1--20, 2024.

\bibitem{DIANA}
J.~Wang, W.~Feng, M.~Huang, S.~Feng, and D.~Du, ``{Improvement of Practical Byzantine Fault Tolerance Consensus Algorithm Based on DIANA in Intellectual Property Environment Transactions},'' \emph{Electronics}, vol.~13, no.~9, p. 1634, 2024.

\bibitem{synchotstuff}
I.~Abraham, D.~Malkhi, K.~Nayak, L.~Ren, and M.~Yin, ``{Sync hotstuff: Simple and practical synchronous state machine replication},'' in \emph{2020 IEEE Symposium on Security and Privacy (SP)}.\hskip 1em plus 0.5em minus 0.4em\relax IEEE, 2020, pp. 106--118.

\bibitem{hotstuff-2}
D.~Malkhi and K.~Nayak, ``{Hotstuff-2: Optimal two-phase responsive bft},'' \emph{Cryptology ePrint Archive}, 2023.

\bibitem{oneshot}
J.~Decouchant, D.~Kozhaya, V.~Rahli, and J.~Yu, ``{OneShot: View-Adapting Streamlined BFT Protocols with Trusted Execution Environments},'' in \emph{IPDPS 2024}, 2024.

\bibitem{flash}
C.~Berger, L.~Rodrigues, H.~P. Reiser, V.~Cogo, and A.~Bessani, ``{Chasing the speed of light: Low-latency planetary-scale adaptive Byzantine consensus},'' \emph{CoRR}, vol. abs/2305.15000, 2023.

\bibitem{reiter1998byzantine}
M.~Reiter and D.~Malkhi, ``{Byzantine quorum systems},'' \emph{Distributed Computing}, vol.~11, no.~4, pp. 203--213, 1998.

\bibitem{round-robin}
D.~Yaga, P.~Mell, N.~Roby, and K.~Scarfone, \emph{{Blockchain Technology Overview}}.\hskip 1em plus 0.5em minus 0.4em\relax NIST Interagency/Internal Report (NISTIR), National Institute of Standards and Technology, Gaithersburg, MD, 2018.

\bibitem{cloudping}
\BIBentryALTinterwordspacing
``{CloudPing},'' Website, {Accessed on June 2, 2024}. [Online]. Available: \url{https://www.cloudping.co/grid/latency/timeframe/1D}
\BIBentrySTDinterwordspacing

\bibitem{codeofconduct}
\BIBentryALTinterwordspacing
{Netherlands Organisation for Scientific Research (NWO)}, ``{Netherlands Code of Conduct for Research Integrity},'' {Accessed on June 19, 2024}. [Online]. Available: \url{https://www.nwo.nl/en/netherlands-code-conduct-research-integrity}
\BIBentrySTDinterwordspacing

\bibitem{gitlab-repo}
\BIBentryALTinterwordspacing
D.~Micloiu, ``{Using Weighted Voting to Optimise Streamlined Blockchain Consensus Algorithms},'' 2024, {Accessed on June 23, 2024}. [Online]. Available: \url{https://github.com/dmicloiu/weightedhotstuff.git}
\BIBentrySTDinterwordspacing

\end{thebibliography}

\clearpage

\pagebreak

\appendix

\section{Appendix}\label{appendix}

\subsection{Hotstuff communication}\label{appendix1}
\begin{figure}[h]
    \centering
    \includegraphics[width=0.5\textwidth]{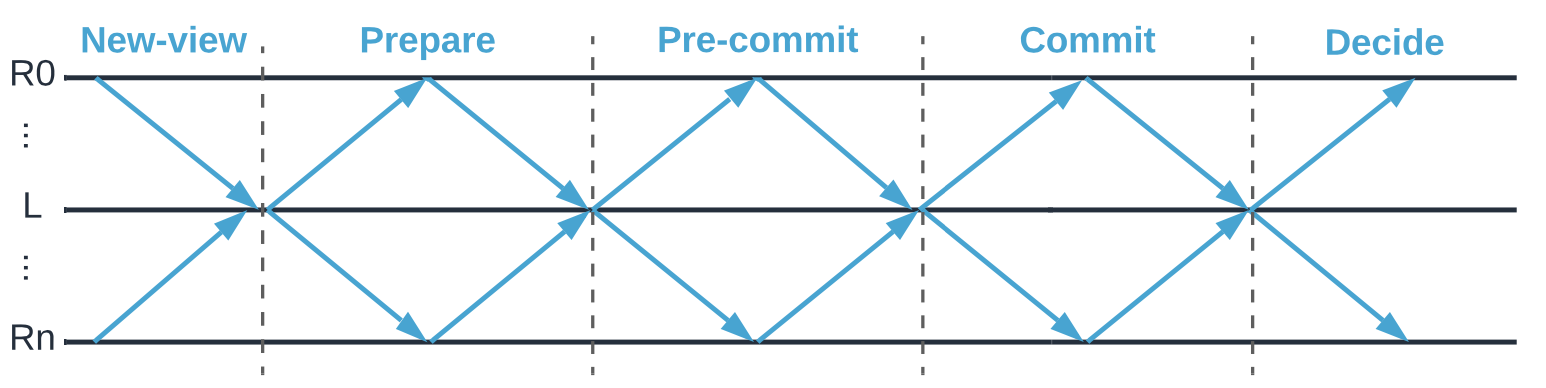}
    \caption{Hotstuff communication phases.}
    \label{fig:comm phases}
\end{figure}

~\cref{fig:comm phases} depicts the five communication phases of Hotstuff~\cite{HotStuff} that together form one \textit{view}. This view has $n$ participating replicas, out of which one is the leader. As the illustration highlights, only the leader communicates with the other replicas, thus achieving linear communication rather than the quadratic one of PBFT~\cite{PBFT}.

\subsection{Continuous Weighted Hotstuff Analysis}\label{appendix2}

\begin{figure}[h]
    \centering
    \includegraphics[width=0.5\textwidth]{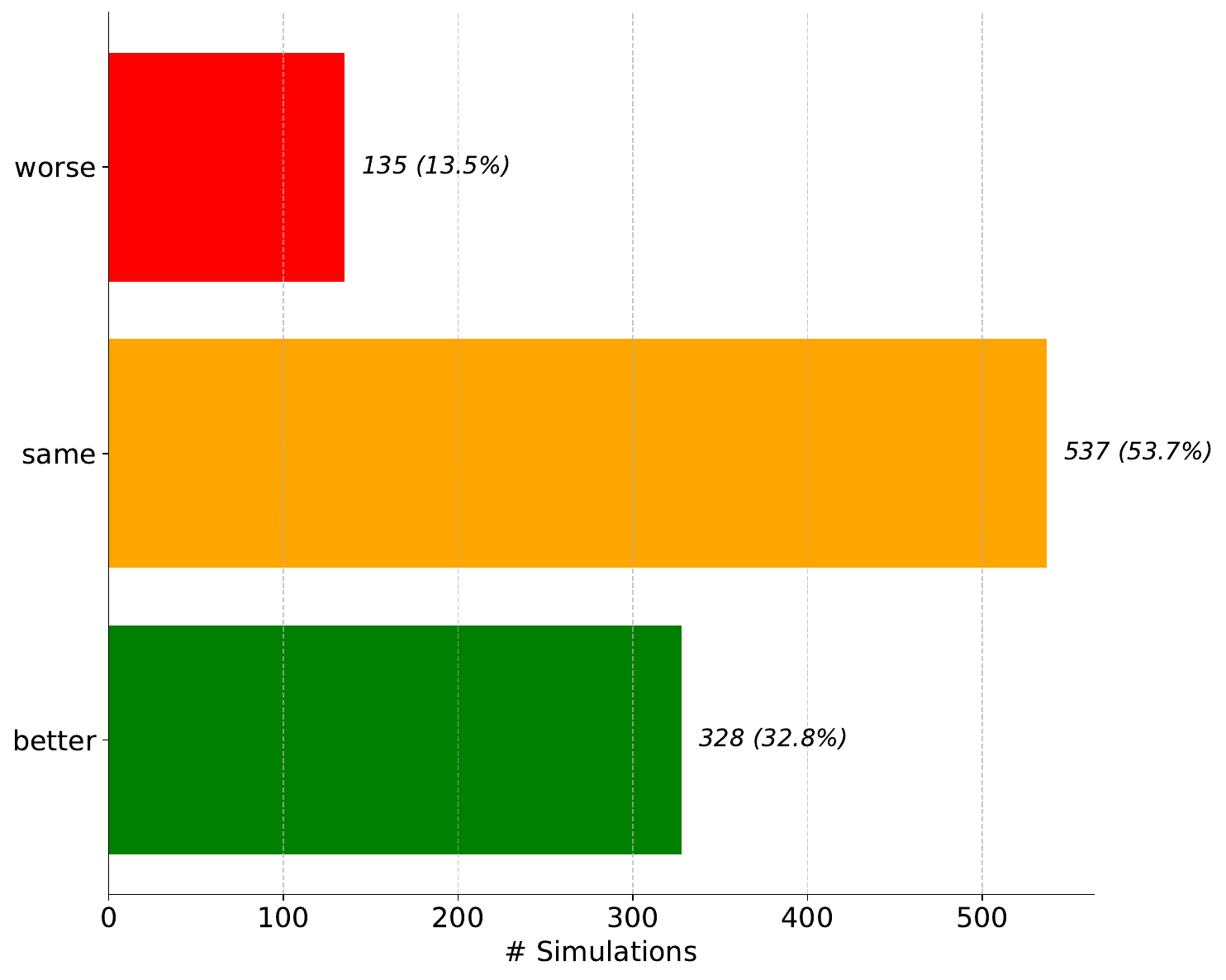}
    \caption{Latency performance comparison between Continuous and Best Assigned Weighted Hotstuff for 1000 \textit{faulty scenario} simulations, $f = 1, \Delta = 1$, $10$ views executed.}
    \label{fig:continuous comparison percentages}
\end{figure}

Compared with the Continuous version, Best Assigned Weighted Hotstuff has higher latency in $13\%$ of the simulations performed (see~\cref{fig:continuous comparison percentages}). Even though the algorithms have matching performance in around $50\%$ of the cases, the generalised weighting scheme still recovers faster in $30\%$. Since the optimisation model we developed relies on \textit{Simulated Annealing}~\cite{baeldungSimulatedAnnealing}, the algorithm can be further tweaked to improve these results. By increasing the hyperparameter for step convergence, we can let the annealing process explore more, eventually reaching the state of a better-performing solution than the Best Assigned one. 

\begin{figure}[h]
    \centering
    \includegraphics[width=0.5\textwidth]{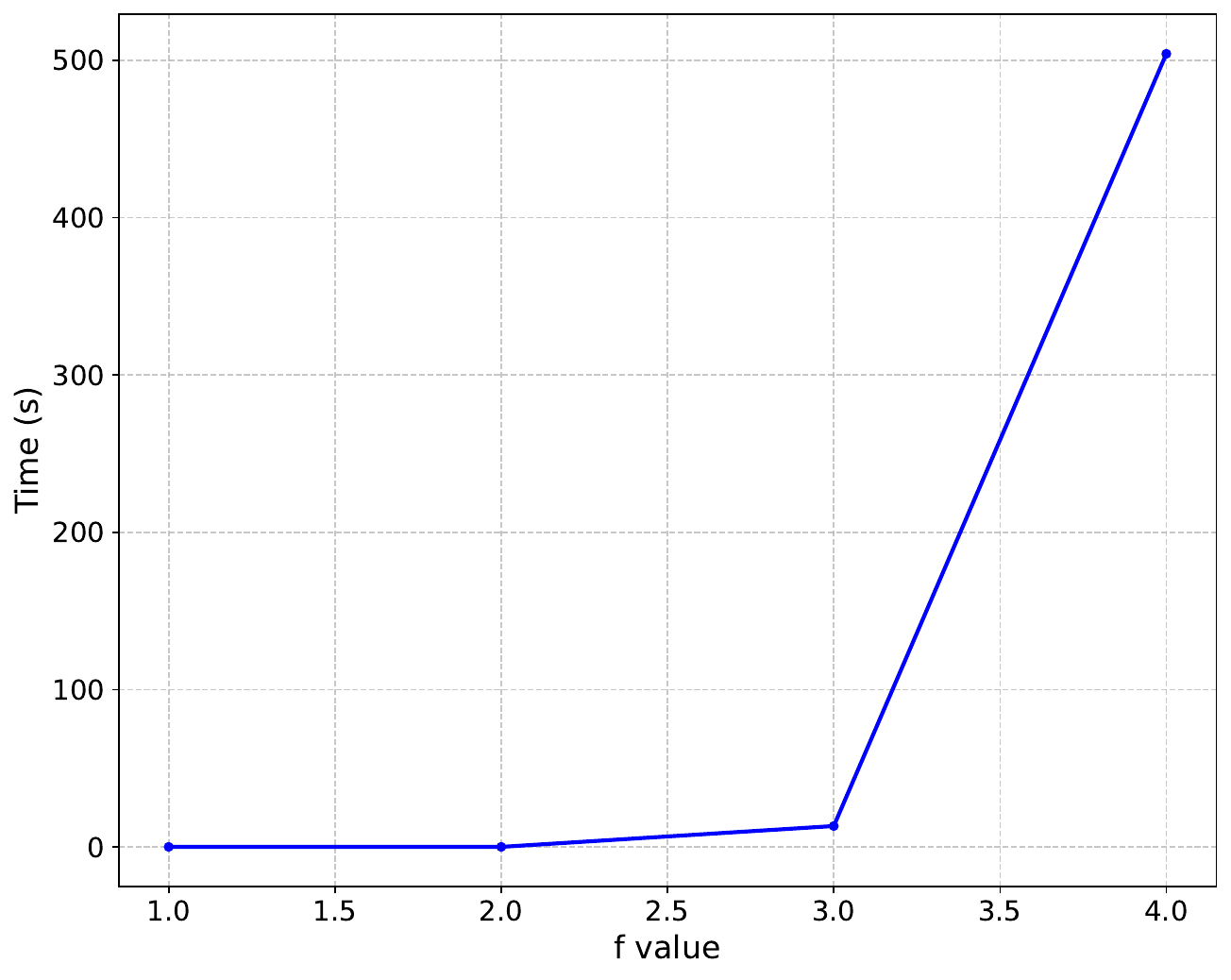}
    \caption{Simulated Annealing convergence time for Continuous Weighted Hotstuff for multiple $f$ values, $1$ view executed.}
    \label{fig:continuous times}
\end{figure}

However, the prediction model we developed in this research has one considerable limitation: \textit{high computational complexity}. Due to the rigorous process of ensuring quorum system safety, the time it takes the \textit{Simulated Annealing} algorithm to converge increases significantly with the number of faulty replicas the system can withstand. In this sense,~\cref{fig:continuous times} showcases that running the prediction model for $f = 4$ took almost $40$ times more than for the precedent value.

\end{document}